# Two-dimensional Keldysh theory for non-resonant strong-field ionization of monolayer 2D materials

Tsing-Hua Her[1†*], Che-Hao Chang[2†], Kenan Darden[1], Tsun-Hsu Chang[2], Hsin-Yu Yao[3]

[1]Department of Physics and Optical Science, The University of North Caroline at Charlotte, Charlotte, NC 28223, USA

[2]Department of Physics, National Tsing Hua University, Hsinchu 300044, Taiwan

[3]Department of Physics, National Chung Cheng University, Chaiyi 621301, Taiwan

[†] These authors contributed equally to this work.

[*] ther@charlotte.edu

## Abstract

The Keldysh theory of photoionization for solids is generalized to atomically thin two-dimensional semiconductors. We derive a closed-form formula and its asymptotic forms for a two-band model with a Kane dispersion. These formulas exhibit characteristically different behaviors from their bulk counterparts which are attributed to the scaling of the 2D density of states. We validate our formulas by comparing them to recent strong-field ionization experiments in monolayer $MoS_2$ with good agreement. Our work is expected to find a wide range of applications in intense light - 2D material interaction.

# Introduction

In bulk solids, strong-field ionization refers to interband transition of electrons induced by light with photon energy smaller than the bandgap energy. It is foundational to a diverse array of light-matter interaction phenomena in solids, including optical-field-driven tunneling [1], terahertz generation [2], low- and high-order harmonic generation [3,4], multiphoton absorption [5], optical injection of spin and charge currents [6,7], nonlinear pulse propagation [8], and laser-induced dielectric breakdown [9]. Theories of strong-field ionization in bulk materials have been developed since the 60s, both perturbative [5] and non-perturbative [10], which provide theoretical frameworks to significantly aid our understanding of the above phenomena. Over the past decade, similar phenomena have been observed in monolayer transition metal dichalcogenides (TMDs) [11-18], but theoretical understanding of strong-field ionization in these materials is far from satisfactory. Experimentally, bulk-equivalent 2- (2PA) and 3-photon absorption (3PA) coefficients of monolayer $MoS_2$ have been measured using different techniques, with data spanning over 3 orders of magnitude [13,19-21]. Theoretically, Zhou et al. have calculated these coefficients using 3D perturbation theory considering various excitonic bound states as the intermediate and final states [20,21]. Although their model can fit experimental data within an order of magnitude by invoking the linewidth of intermediate and final excitonic states as a fitting parameter, the justification for using different linewidths for 2PA and 3PA measurements, even for the same sample, is not clear [20,22]. On the other hand, although strong-field-driven electron tunneling in monolayer 2D materials has been demonstrated in photoelectron emission [23,24], interband transition within these materials via tunneling has not been reported. The latter process is, however, assumed to play a critical role in initiating high-order harmonic generation (HHG) in monolayer $MoS_2$ [11]. The authors measured HHG yields from the monolayer and a single layer in the bulk material, and calculated their theoretical strength based on semiclassical equations of motion in a single particle band. Assuming the same initial electron density in the monolayer and the single layer of the bulk, their ratios of HHG efficiency obtained experimentally and theoretically are off by one order of magnitude, and this discrepancy was not explained. All of the above discrepancies between experiments and modeling highlight the lack of theoretical understanding of strong-field ionization in monolayer 2D materials.

In this paper, we report a new formalism of strong-field ionization for atomically thin two-dimensional semiconductors based on 2D Keldysh (KLD) theory. We take this approach because the original Keldysh theory, which is non-perturbative in nature, has been shown to provide a uniform description of multiphoton and tunneling ionization in atoms and in bulk solids [10]. With a simplified band dispersion, Keldysh presented analytical formulas for the cycle-averaged non-resonant ionization rate in atoms and bulk solids induced by a monochromatic electric field of arbitrary strength [10]. Even though Keldysh's formulas are known to have limitations [25-29], they are widely employed for qualitative modeling of strong-field ionization in bulk solids because of their analyticity. In this report, we show that 2D Keldysh formulas bear the same simplicity as the ones for bulk solids. We validate our theory by comparing it to recent experiments and modeling of strong-field ionization in monolayer TMDs with good agreement.

# 2D KLD formulism

Here we consider a direct above-bandgap transition in monolayer 2D materials excited with a normally incident linearly polarized light. Although monolayer 2D materials are known to exhibit robust valley-spin polarization by the helicity of the optical pumping, such selection rules were shown to only be valid for on-resonance excitation of A excitons near K and -K valleys in the momentum space [30]. For above bandgap transitions in monolayer MoS2, Ref. [30] demonstrated experimentally that the above selection rules are relaxed, in that both left and right circularly polarized lights at 2.33 eV can simultaneously populate K and -K valleys with equal probability. Given that its typical quasi-particle bandgap is around 2.4 eV [31], this observation suggests that above-bandgap transitions in monolayer MoS2 can be directly induced by linearly polarized light for both valleys.

We denote the electric field inside the monolayer as $\mathbf{F}(t) = \mathbf{F}\cos(\omega t)$, where $\mathbf{F}$ represents the electric field vector and $\omega$ is the field angular frequency. Under a two-band model, the interband transition of an electron from the valence to the conduction band creates an electron-hole pair whose energy is $\varepsilon(\mathbf{p}) = \varepsilon_c(\mathbf{p}) - \varepsilon_v(\mathbf{p})$, where $\varepsilon_{c,v}(\mathbf{p})$ are the energies of the corresponding electron and hole and $\mathbf{p}$ is the relative crystal momentum $\mathbf{p}$ measured from critical points. The surface ionization probability for the monolayer 2D materials can be written following Keldysh [32] as

$$w = \frac{2\pi}{\hbar} \int \frac{d^2 p}{(2\pi\hbar)^2} |L_{cv}(\mathbf{p})|^2 \sum_n \delta(\bar{\varepsilon}(\mathbf{p}) - n\hbar\omega), \tag{1}$$

where the quasienergy $\bar{\varepsilon}(\mathbf{p}) = \frac{1}{2\pi}\int_{-\pi}^{\pi} \varepsilon(\mathbf{p} + \frac{e\mathbf{F}}{\omega}\sin x)dx$ and the ionization amplitude $L_{cv}(\mathbf{p})$ is defined by [32]

$$L_{cv}(\mathbf{p}) = \frac{1}{2\pi} \oint_{C_{in}} V_{cv}\left(\mathbf{p} + \frac{e\mathbf{F}}{\omega}u\right) \exp\left[\frac{i}{\hbar\omega}\int_0^u \varepsilon\left(\mathbf{p} + \frac{e\mathbf{F}}{\omega}v\right)\frac{dv}{\sqrt{1-v^2}}\right] du, \tag{2}$$

where $V_{cv}(\mathbf{p}) = i\hbar\int u_{\mathbf{p}}^{c*}(\mathbf{r}) e\mathbf{F}\cdot\nabla_{\mathbf{p}} u_{\mathbf{p}}^{v}(\mathbf{r})d\mathbf{r}$ is the optical matrix element and $u_{\mathbf{p}}^{c,v}(\mathbf{r})$ are periodic functions with the translational symmetry of the lattice. Following Keldysh, we calculate Eq. (2) using the saddle point method with a contour that encloses a branch cut along [-1, 1], and adopt a Kane band dispersion model, $\varepsilon(\mathbf{p}) = \Delta(1 + \mathbf{p}^2/\Delta m)^{1/2}$, where $m$ is the reduced mass of the electron-hole pair and $\Delta$ is the quasiparticle bandgap. The expression of $\bar{\varepsilon}(\mathbf{p})$ is identical to Eq. (35) in Ref. [32], and $L_{cv}(\mathbf{p})$ can be written as [32,33]

$$L_{cv}(\mathbf{p}) = \frac{\hbar\omega}{3}\left[\exp(\varphi_0 + i\varphi) - \exp(\varphi_0 - i\varphi)\right] \tag{3}$$

with

$$\begin{cases} \varphi_0 = -\frac{\Delta}{\hbar\omega}\left(\frac{1}{\gamma_2}[K_2 - E_2] + \frac{x^2\gamma_2}{2}[K_2 - E_2] + \frac{y^2\gamma_2}{2}K_2\right) \\ \varphi = -x\frac{\Delta}{\hbar\omega}\arctan(\gamma) \end{cases}, \tag{4}$$

where $x = p_\parallel/\sqrt{m\Delta}$, $y = p_\perp/\sqrt{m\Delta}$, and $p_\parallel$ and $p_\perp$ are the components of the crystal momentum parallel and perpendicular to the electric field $\mathbf{F}$, respectively. Moreover, $K_{1,2} = K(\gamma_{1,2})$ and $E_{1,2} = E(\gamma_{1,2})$ are the complete elliptic integrals of the first and second kind of $\gamma_1 = (1 + \gamma^2)^{-1/2}$ and $\gamma_2 = \gamma(1 + \gamma^2)^{-1/2}$, where $\gamma = \omega\sqrt{m\Delta}/(e\mathbf{F})$ is the Keldysh parameter. Eq.(1) can then be reduced to yield the photoionization rate for 2D materials

$$w = \frac{\omega\pi}{9}\left(\frac{m\omega}{\gamma_2\hbar}\right)\sqrt{\frac{1}{E_1 K_1}} Q(\gamma,\tilde{n}) \exp\left(-\pi N \frac{K_2 - E_2}{E_1}\right), \tag{5}$$

where $N = <\tilde{n}+1>$ refers to the integer part of $\tilde{n}+1$, $\tilde{n} = \tilde{\Delta}/\hbar\omega$ is effective bandgap energy $\tilde{\Delta} = 2\Delta E_1/\pi\gamma_2$ normalized by the photon energy $\hbar\omega$, and

$$Q(\gamma,\tilde{n}) = \sum_{n=0}^{\infty} \exp\left(-\pi n \frac{K_2 - E_2}{E_1}\right) \cdot \exp\left[-\frac{\pi^2}{4 E_1 K_1}(N - \tilde{n} + n)\right] \cdot I_0\left[\frac{\pi^2}{4 E_1 K_1}(N - \tilde{n} + n)\right]. \tag{6}$$

Compared to the 3D Keldysh formula, besides the pre-factors, the major difference is that the Dawson integral $\Phi$ in Eq. (39) of Ref. [32] is replaced by an exponential term multiplied by a $0^{th}$ order modified Bessel function $I_0$.

In the limit of low frequencies and strong fields, i.e. when $\gamma \ll 1$ and tunneling ionization dominates, Eq. (5) reduces to

$$w_{TI} \approx \frac{2^{\frac{3}{2}}}{9\pi} \frac{\Delta}{\hbar}\left(\frac{m\Delta}{\hbar^2}\right) \left(\frac{\hbar e F}{m^{\frac{1}{2}}\Delta^{\frac{3}{2}}}\right)^2 \exp\left[-\frac{\pi}{2}\frac{\sqrt{m\Delta^3}}{\hbar e F}\left(1 - \frac{\gamma^2}{8}\right)\right]. \tag{7}$$

In the opposite limit of high frequencies and low fields, i.e. when $\gamma \gg 1$, the first term in Eq.(6) decays dramatically with respect to $n$. In this regime, the effective bandgap reduces to $\tilde{\Delta} \approx \Delta + e^2 F^2/4m\omega^2$ and Eq. (5) becomes the following multiphoton ionization expression of order $N$

$$w_{MPI} \approx \frac{2\omega}{9}\left(\frac{m\omega}{\hbar}\right) \exp\left[\left(1 - \frac{1}{2\gamma^2}\right)N + \tilde{n}\right] I_0[N - \tilde{n}]\left(\frac{1}{16\gamma^2}\right)^N. \tag{8}$$

The full expression and its asymptotic forms are plotted in Figure 1 for 800 nm light incident on a hypothetical monolayer 2D material with a direct quasiparticle bandgap of 2.4 eV and a reduced mass $m = 0.215\, m_e$, where $m_e$ is the free electron mass. The TI (black dash) and MPI (green dash) limiting curves work well and their intersection delineates the transition between of MPI and TI dominated regimes, which occurs at F = $4.75 \times 10^9$ V/m ($\gamma$ = 0.85) and coincides with the first channel closure in this case. For comparison, Figure 1 also plots the 3D KLD rate multiplied by a monolayer thickness of 0.63 nm to match the dimensionality of the 2D rate. The 2D rate has a similar trend to the 3D rate but exceeds it in the low-field (MPI) regime and is eventually overtaken in the high-field (TI) regime. The field at which the 3D and 2D rates cross can be found to be $F_{cross} = 2\pi^2 \hbar\sqrt{\Delta}/(d^2 e\sqrt{m})$, which is ~ $4.3 \times 10^{10}$ V/m ($\gamma_{cross}$ ~ 0.088) for this case, given $d$ is the monolayer thickness. The 2D rate also features relatively abrupt channel closure when compared to the 3D rate. This contrast is perhaps more evident in the inset which plots their ratio. This ratio is nearly constant ~ 3 at low fields and decays as the field increases. On top of this general trend, there are spikes at channel closures, which attenuate in strength as the field increases. We attribute these behaviors to the different scaling behaviors of densities of states in 2 and 3D [34], but a thorough understanding deserves future investigation.

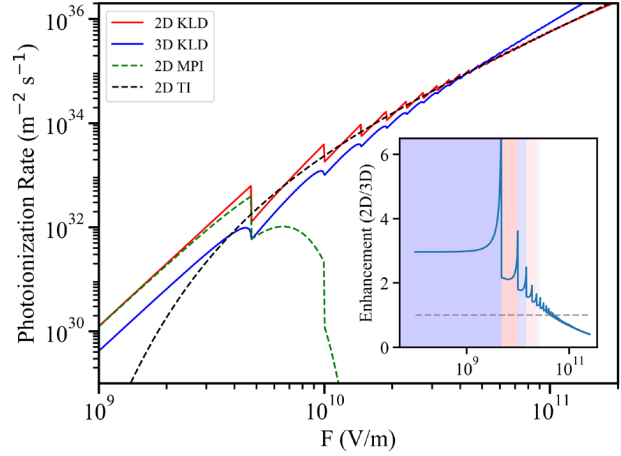

Figure 1: Plot of the 2D photoionization rate (red, Eq. (5)) and its asymptotic approximations, MPI (green, Eq. (8)) and TI (black, Eq. (7)) as functions of the incident electric field. Keldysh's 3D rate normalized by a material width of 0.63 nm is plotted for comparison. All curves were generated with a material bandgap of 2.4 eV, an effective mass of 0.215 electron masses, and an incident field with an 800 nm wavelength and without accounting for spin degeneracy.

A complimentary view of this comparison is seen by plotting these two rates as a function of photon energy normalized by the bandgap in Figure 2 for a field strength of $10^7$ V/m. In this MPI-limiting regime, $\tilde{\Delta} \approx \Delta$ and the channel closure events can be more directly associated with transitions between $N$ and $N+1$ multiphoton orders at $\hbar\omega/\Delta = N^{-1}$. Again, the inset shows the ratios of the 2D and 3D KLD rates. Interestingly, it features significantly stronger enhancement for all channel closure events, compared to the inset of Figure 2. This can be understood as follows: near the band edge where the photon energy detuning $\delta_N = N\hbar\omega - \Delta$ approaches zero, the 2D MPI rate $w_{MPI}^{2D} \propto I_0(\delta_N) \propto 1$, whereas the 3D MPI rate $w_{MPI}^{3D} \propto \Phi(\delta_N) \propto \sqrt{\delta_N}$ [5,33]. Such a photon energy scaling is consistent with that of density of states $g(\varepsilon)$ for Kane dispersion, where $g^{2D}(\varepsilon) \propto \varepsilon/\Delta$ and $g^{3D}(\varepsilon) \propto \varepsilon\sqrt{(\varepsilon^2 - \Delta^2)/\Delta}/\Delta$. The ratio of these rates at the band edge is therefore proportional to $\delta_N^{-0.5}$, which is singular right at the band edge for all photon orders.

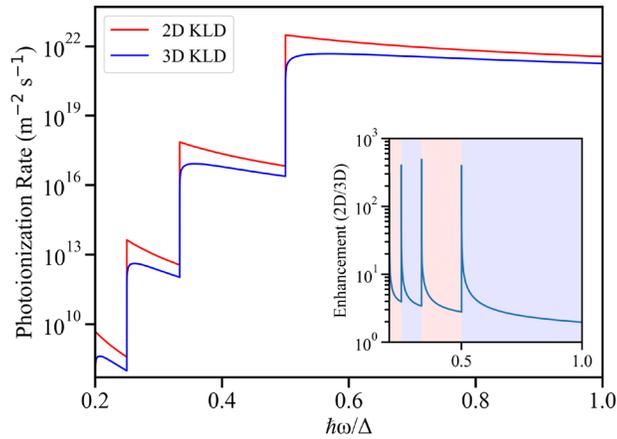

Figure 2: Photoionization rates for a monolayer calculated using 2D KLD (solid red) formula, and a single layer of a bulk crystal using 3D KLD (solid blue) formula, as a function of photon energy for a constant field strength of $10^7$ V/m. All materials assume a bandgap energy of 2.4 eV and a layer thickness of 0.63 nm. The inset shows the ratio of the 2D and 3D KLD rates.

In the multiphoton regime, we can define an internal surface $N$-photon ionization cross section $\sigma_N$ (in units of (length)$^{2N-2}$/(power)$^N$/$time$) using Eq. (8) by $w_N = \sigma_N I^N$ [35] and an internal surface $N$-photon absorption coefficient $\alpha_N$ (in units of (length)$^{2N-2}$/(power)$^{N-1}$) by $\Delta I = w_N \cdot N\hbar\omega = \alpha_N I^N$, where $I = \frac{1}{2}c\varepsilon_o n_{2DM} F^2$ is the internal intensity inside the monolayer. Analytical expressions for $\sigma_N$ and $\alpha_N$ can be found in Section 1 of Supplemental Material (SM). The use of internal quantities ensures that they are genuine properties of the monolayer and independent of the supporting substrates. Previously we have demonstrated that the external ablation threshold fluence of monolayer MoS$_2$, referenced to the incident fluence in the air, is substrate dependent [15]. More specifically, it is inversely proportional to the square of the electric field on the surface of the supporting substrate, which results from the interference between the incident and the reflected field from the substrate [15]. Conversion between the external and internal nonlinear absorption coefficients can be found in Section 2 of the SM.

## Comparison with 2PA & 3PA exp. & theory of Refs. [20,21]

To validate our findings, we apply them to literature data on multiphoton absorption of monolayer MoS$_2$ [13,19-21]. The most complete experimental data was collected by Zhou et. al., who recorded bulk-equivalent 2- and 3-photon absorption coefficients of mechanically exfoliated monolayer MoS$_2$ on 285-nm-thick SiO$_2$/Si over a wide range of wavelengths using a photoconductivity technique [20,21]. This method is superior to intensity or Z-scan techniques [13,19] which have difficulty excluding other photon depletion processes, including Kerr harmonics [36], low-order injection harmonics [4], free carrier absorption, and substrate absorption. In addition, this method deduced their results by referencing to one photon absorption to bypass uncertainties in carrier lifetime and mobility [20,21]. Refs. [20,21], however, reported external coefficients referenced to the incident intensity, which we converted to internal surface multiphoton absorption coefficients by a procedure described in Section 3 of the SM. These are reproduced in Figure 3 (black dots). Refs. [20,21] also reported theoretical bulk-equivalent 2- and 3-photon absorption coefficients using 3D perturbation theory with 1$s$- and/or low-lying $np$-excitons as intermediate states and superpositions of high lying excitons as final states, which we converted to corresponding surface values by multiplying by the thickness of the monolayer (Figure 3 blue dashed curve). Finally, the red solid curves in Figure 3 are our theoretical predictions (Eq. (8)) for monolayer MoS$_2$, employing two valence bands (a bandgap $\Delta = 2.4$ eV [31] separated by a split-off energy $\Delta_{so} = 0.15$ eV [37]), a reduced mass $m = 0.215\, m_e$ [38], and a 2× degeneracy for K and -K valleys [30]. The second humps away from the band edge in these curves are the contributions of the lower valence band.

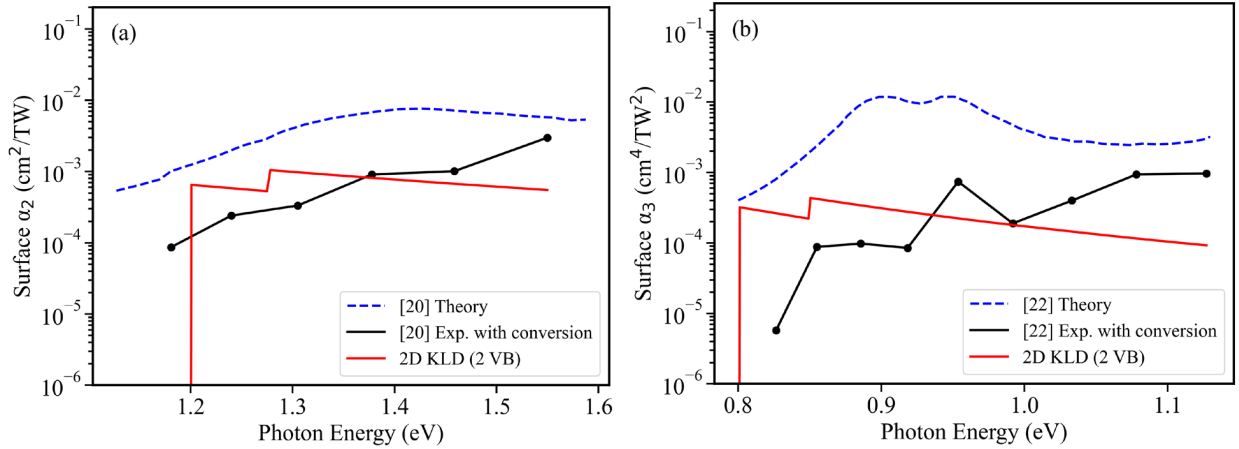

Figure 3: Internal surface 2PA absorption coefficient $\alpha_2$ (a) and surface 3PA absorption coefficient $\alpha_3$ (b) for monolayer $MoS_2$. Black dots are data converted from Ref. [20,21], and blue dash curves are their theoretical prediction based on 3D perturbation theory involving excitonic bound states. Both are converted to internal surface values following a procedure described in Section 2 of the SM. Red solid curves are our theoretical prediction based on 2D KLD formula.

After this conversion, the predictions of their theory exceed their 2PA experimental data by nearly one order of magnitude across the majority of the spectra and their 3PA data by as much as two orders of magnitude. This could be due to two reasons. Firstly, they treated the linewidths of high-lying excitons as a fitting parameter, yielding a linewidth ~ 0.46 eV for 2PA [20] and 0.15 eV for 3PA [21] for the same sample. Such an inconsistency was not explained, undermining their model's validity. Secondly, their use of the Lorentz local field correction is questionable, as it is valid for highly localized electrons in relatively distant atoms in a solid crystal with a cubic symmetry [39], whereas $MoS_2$ monolayer does not have a cubic symmetry, and its free excitons are delocalized. The predictions of our MPI formula, on the other hand, underestimate (overestimate) the experimental data at higher (lower) photon energies by no more than one order of magnitude, except very close to the band edge. Excluding these, the magnitude of the discrepancies is within the range reported for similar comparisons in bulk solids [5,28,29,40,41]. The accuracy of our 2D formula can be potentially improved in several ways by borrowing strategies developed for 3D KLD. Firstly, including the interference of the two saddle points in the approximation of the ionization amplitude $L_{cv}(\mathbf{p})$ has been shown to reduce 3D KLD two-photon prediction near the band edge [33]. Secondly, numerical evaluation of Keldysh's equations (Eqs. (27)-(30) in Ref. [32]) has been shown to reduce overestimation (underestimation) of 3D KLD two- and three-photon predictions near (away from) the band edge by avoiding the saddle point and small momentum approximations used in the traditional evaluation of the ionization amplitude $L_{cv}(\mathbf{p})$ [28]. Lastly, using more realistic band structures for the materials of interest to replace Kane or parabolic dispersion has been shown to reduce underestimation [5].

## Comparison with HHG exp. & theory of Ref. [11]

As a second test of our 2D Keldysh theory, we apply it to tunneling ionization in monolayer $MoS_2$. Although tunneling-induced interband transitions within monolayer 2D materials have not been directly measured, they are assumed to be responsible for seeding high-order harmonic generation (HHG) within these materials [11]. According to Liu et al., a small fraction of the electrons was assumed to tunnel from the valence band to the conduction band near the direct gap at the peak of the pump field. The electrons and holes are subsequently accelerated in their respective bands by the driving pulse. The nonlinear currents resulting from such intraband motion lead to HHG. Based on their experimental conditions (F = 3.3 V/nm, Δ = 2.4 eV, photon energy $\hbar\omega$ = 0.3 eV, reduced mass of electron-hole pair $m$ = 0.215 $m_e$), the Keldysh parameter $\gamma$ is ≈ 0.24, confirming that the photoionization in their experiment is in the tunneling regime. In their work, Liu et al. compared the odd-order HHG yields from a $MoS_2$ monolayer and a single layer of a 60-nm-thick $MoS_2$ bulk crystal. They also calculated the theoretical yields of these two configurations based on semiclassical equations of motion in a single particle band including band dispersion, assuming the same initial electron density. Figure 4 shows these two ratios, obtained experimentally (blue dots) and theoretically (green dots), as a function of the odd harmonic order. Except for the 9th order, both ratios increase monotonically with increasing harmonic order, while the theoretical ratio is smaller than its experimental counterpart by nearly one order of magnitude. As it has been experimentally demonstrated that tunneling ionization rate and HHG efficiency are strongly correlated in bulk $SiO_2$ [42], Ref. [11]'s assumption of the same initial electron density in these two materials is questionable. We therefore calculate tunneling ionization rate for a $MoS_2$ monolayer $w_{TI}^{2D}$ using Eq. (7) and that for a single layer in the bulk $MoS_2$ $w_{TI}^{3D}$ using the 3D Keldysh equation (Eq. (37) in Ref. [32]) multiplied by the monolayer thickness of 0.63 nm and by a factor of 2 to account for degeneracy of the conduction bands in the bulk crystal [30]. This translates to a density ratio ~ 6.5, indicating the initial tunneling electron density for the same driven field is substantially higher in the 2D monolayer due to the enhanced density of states associated with the reduced dimensionality. Incorporating this ratio into Ref. [11]'s theoretical ratio (green dots), we obtain the red dots in Figure 4, which are in good agreement with their experimental ratio to within a factor of 2 for the harmonic orders 7, 11, and 13. A large discrepancy remains for the 9th order due to reasons beyond the difference in the tunneling rates. We want to emphasize that such a good agreement is robust against the qualitative nature of the KLD formulas as taking the ratio removes common errors in $w_{TI}^{2D}$ and $w_{TI}^{3D}$ introduced from analytical approximation of the ionization amplitude $L_{cv}(\mathbf{p})$ in the Keldysh theory [28].

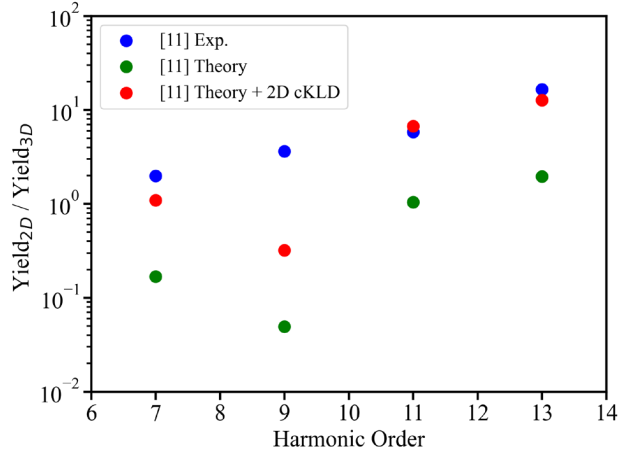

Figure 4: The ratio of HHG yields in monolayer $MoS_2$ and a single layer in bulk $MoS_2$ as a function of the odd harmonic order obtained experimentally (blue) [11], theoretically considering only the intraband motion of conduction electrons (green) [11], and theoretically combining Ref. [11] and tunneling electron density ratio predicted by our 2D KLD theory (red).

# Conclusion

In conclusion, the Keldysh theory of photoionization for bulk dielectrics is generalized to atomically thin two-dimensional semiconductors. We derive a closed-form photoionization formula and its asymptotic forms for a two-band model with a Kane dispersion. Compared to its bulk counterpart, the 2D KLD rate is enhanced in the multiphoton limit and suppressed in the tunneling limit. In the intermediate regime, it displays more abrupt channel closure events, whose strength attenuates towards the tunneling limit. These phenomena are consistent with the scaling of the electronic density of states in reduced dimensionality. Our theory is validated against recent strong-field ionization experiments in monolayer $MoS_2$. Firstly, our MPI formula simultaneously reproduces the general trends of experimental 2PA and 3PA absorption spectra without the use of fitting parameters. The magnitude of the discrepancies from these experimental data is within that observed for similar 3D KLD predictions. Strategies for improving the accuracy of the 2D MPI formula are proposed. Secondly, our theory successfully explains the discrepancy between experiment and modeling of the HHG efficiency ratio between a monolayer and a single layer in bulk $MoS_2$ by predicting quantitatively the difference in their tunneling ionization rates. Considering the tremendous success of the original atomic and solid Keldysh theories in describing strong-field optical phenomena, our theory is expected to find a wide range of applications in intense light-2D material interaction.

# Acknowledgements

H.-Y Yao acknowledges the support from the National Science and Technology Council of Taiwan (NSTC 112-2112-M-194-006-MY3).

# Supplemental Materials

## Section 1: Internal multiphoton ionization cross section $\sigma_N$ and multiphoton absorption coefficient $\alpha_N$

To compare to microscopic theory of optical properties, the electromagnetic field inside the material should be invoked. We therefore define an internal 2D $N$-photon ionization cross section in terms of the internal intensity $I = \frac{1}{2}c\varepsilon_o n_{2DM} F^2$ by $w_N = \sigma_N I^N$ [1]. We note that this definition of ionization cross section is different from others [2,3]. From Eq. (8) (the MPI limit) of the main text, $\sigma_N$ can be written as

$$\sigma_N = \frac{2m\omega^2}{9\hbar} \left( \frac{1}{8\gamma^2 c\varepsilon_o n_{2DM} F^2} \right)^N \exp\left[ \left(1 - \frac{1}{2\gamma^2}\right) N + \tilde{n} \right] I_0 [N - \tilde{n}]. \quad (S1)$$

Similarly, we can define an internal 2D $N$-photon absorption coefficient $\alpha_N$ in terms of the internal intensity by $\Delta I_N = w_N \cdot N\hbar\omega = \alpha_N I^N$. It is easy to see that $\alpha_N = \sigma_N \cdot N\hbar\omega$ and can be written as

$$\alpha_N = \frac{2Nm\omega^3}{9} \left( \frac{1}{8\gamma^2 c\varepsilon_o n_{2DM} F^2} \right)^N \exp\left[ \left(1 - \frac{1}{2\gamma^2}\right) N + \tilde{n} \right] I_0 [N - \tilde{n}]. \quad (S2)$$

## Section 2: External-to-internal conversion of multiphoton absorption coefficient

It is well known that the light field inside the 2D materials (called the internal field $F$) is different from the incident field (called the external field $F'$) due to interference influenced by the supporting substrate [4]. These two quantities are related according to $F = \eta F'$, where $\eta$ is the field enhancement factor. Analytical expressions of $\eta$ for various substrates can be found in [5]. Such an effect is well known in weak-field optical studies of 2D materials (*e.g.*, optical contrast, photoluminescence and Raman scattering [6] [4], SHG [7], and CW laser thinning [8]), and was recently demonstrated in femtosecond laser ablation of monolayer MoS2, where we showed that the external ablation threshold $F'_{th}$ is substrate dependent and $\xi F'_{th}$ is a constant, where $\xi = \eta^2$ is the incident intensity enhancement factor [5]. It is therefore important to distinguish between internal and external optical properties when studying strong-field physics of 2D materials.

For nonlinear transmission experiments, the bulk-equivalent internal ($\alpha_N$) and external ($\alpha'_N$) $N$-photon absorption coefficients are defined based on experimental observable $dI'/dz$ according to

$$\frac{dI'}{dz} = \alpha'_N I'^N = \alpha_N I^N, \tag{S3}$$

where $I' = \frac{1}{2}c\varepsilon_0 E'^2$ and $I = \frac{1}{2}c\varepsilon_0 n_{2DM} E^2$ are the external (incident) and internal intensity, respectively, and $n_{2DM}$ is the internal refractive index of the 2D material. $\alpha_N$ and $\alpha'_N$ are therefore related by

$$\frac{\alpha'_N}{\alpha_N} = \frac{1}{(n_{2DM}\xi)^N}. \tag{S4}$$

## Section 3: External-to-internal conversion of multiphoton absorption coefficient for Zhou et al [9,10].

Zhou et al. employed a nonlinear photoconductivity technique to measure multiphoton absorption coefficients by referencing to one photon absorption to bypass uncertainties in carrier lifetime and mobility [9,10]. For two-photon absorption ($N = 2$), we have, according to Eq. (S5),

$$\frac{\alpha_2}{\alpha_1} = \frac{\alpha'_2}{\alpha'_1} \frac{n_1 \xi_1}{(n_2 \xi_2)^2}, \tag{S5}$$

where $n_i = n_{2DM}(v_i)$ and $\xi_i = \xi(v_i)$, and $v_i$ is the frequency of the $i^{th}$-photon absorption process. The external quantity $\alpha'_2/\alpha'_1$ is related to the experimental photocurrent ratio $J_2/J_1$ by (see Appendix 1 in [11] for a full derivation) [9]

$$\frac{J_2}{J_1} = \frac{v_1 f_2}{\sqrt{2\pi}v_2 (w_2)^2 t_2} \frac{\alpha'_2}{\alpha'_1} \frac{(E_2)^2}{E_1}, \tag{S6}$$

where $w_i$ is the $1/e^2$-intensity beam radius, $t_i$ is the pulse width, $f_i$ is a geometric factor, and $E_i$ is the pulse energy seen by the monolayer MoS$_2$ for the $i^{th}$-photon absorption process [9]. To extract $\alpha'_2$ from Eq. (S7), Ref. [9] erroneously quoted the internal 1-photon absorption coefficient $\alpha_1$ [12] for $\alpha'_1$. For convenience, we call their extracted value $\alpha''_2$. $\alpha''_2$ and $\alpha'_2$ are therefore related by

$$\frac{J_2}{J_1} \propto \frac{\alpha'_2}{\alpha'_1} = \frac{\alpha_2}{\alpha_1}\frac{(n_2\xi_2)^2}{n_1\xi_1} = \frac{\alpha''_2}{\alpha_1}, \tag{S7}$$

which can be simplified to yield

$$\alpha_2 = \alpha''_2 \frac{n_1 \xi_1}{(n_2 \xi_2)^2}. \tag{S8}$$

For 3-photon absorption ($N = 3$), we can follow similar steps to find

$$\alpha_3 = \alpha''_3 \frac{n_1 \xi_1}{(n_3 \xi_3)^3}. \tag{S9}$$

We note that Eqs. (S9)-(S10) are only valid for Refs. [9,10].

To use Eqs. (S9) and (S10), we need to have the $n_i$ and $\xi_i$ used in Refs. [9,10]. The broadband refractive index $n_i = n(h\nu_i)$ of $MoS_2$ can be extracted from [13] and is shown in Figure S1(a). The incident intensity enhancement factor $\xi_i = \xi(h\nu_i)$ is calculated using Eq.(3) from [5] for the 285-nm-thick $SiO_2$/Si substrate. For both plots, the red and blue dots are for 3PA and 2PA processes, respectively. As shown, $n$ of $MoS_2$ increases gradually whereas $\xi$ decreases drastically with photon energy.

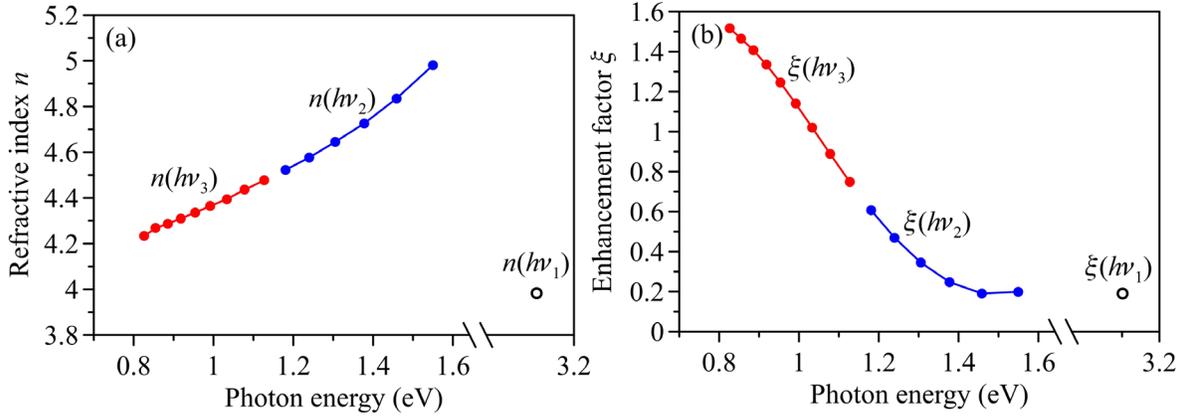

Figure S1: Refractive index $n(h\nu)$ (a) and incident enhancement factor $\xi(h\nu)$ (b) as a function of photon energy used in Refs. [9,10].

Figure S2 plots the conversion factor, the second term on the right-hand side of Eqs. (S9)-(S10), as a function of the photon energy. The conversion factor for 2PA process is in the range of $1 - 10^{-1}$, whereas that for 3PA process is in the range of $10^{-2} - 10^{-3}$. The latter is ~2 orders of magnitude smaller than the former.

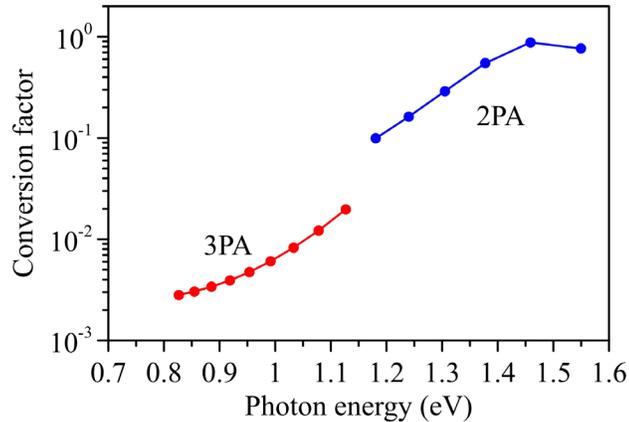

Figure S2: Conversion factors for 2PA (blue) and 3PA (red) as a function of photon energy.

Finally, Figure S3 compares the MPA coefficients before (green circle) and after (black circle) the conversion. It shows that the internal MPA coefficients are generally smaller than those reported in [9] and [10], with a more pronounced reduction for 3PA than 2PA, according to Figure S2.

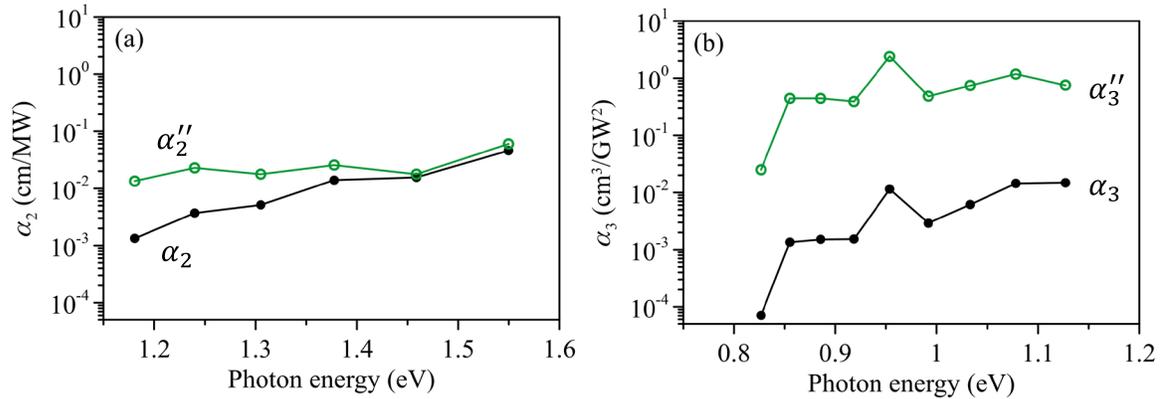

Figure S3: (a) Bulk-equivalent 2PA coefficients before and after conversion ($\alpha_2''$ and $\alpha_2$, respectively). (b) Bulk-equivalent 3PA coefficients before and after conversion ($\alpha_3''$ and $\alpha_3$, respectively). $\alpha_2''$ and $\alpha_3''$ are experimental data directly extracted from Fig. 5 of [9] and Fig. 3(a) of [10], respectively.

We note that the $\alpha_2$ and $\alpha_3$ illustrated in Figure S3 are the bulk-equivalent 2PA and 3PA coefficients for monolayer $MoS_2$. They can be converted into surface MPA coefficients by multiplication with the monolayer thickness of $MoS_2$, the results of which are shown in Figs. 3(a) and 3(b) of the main text.